\documentstyle[preprint,aps]{revtex}
\begin{document}
\draft
\preprint{\vbox{\hbox{IFT--P.025/97}\hbox{hep-ph/9703388}}}
\title{Anomalous Couplings in \boldmath{$e^+ e^- \to W^+ W^-
\gamma$} at LEP2 and NLC}
\author{F.\ de Campos, S.\ M.\ Lietti, S.\ F.\ Novaes and R.\ Rosenfeld}
\address{Instituto de F\'{\i}sica Te\'orica, 
Universidade  Estadual Paulista, \\  
Rua Pamplona 145, CEP 01405--900 S\~ao Paulo, Brazil.}
\date{\today}
\maketitle
\widetext
\begin{abstract}
We present sensitivity limits on the coefficients of a
dimension--6 effective Lagrangian that parametrizes the possible
effects of new physics beyond the Standard Model. Our results are
based on the study of the process $e^+ e^- \to W^+ W^- \gamma$ at
LEP2 and NLC energies. In our calculations, we include {\it all}
the new anomalous interactions, involving vector and Higgs
bosons, and take into account the Standard Model irreducible
background. We analyse the impact of these new interactions on
the total cross section, including the effects of the initial
electron  and final $W$ polarizations. We then focus on the
operators that will not be constrained by the $e^+ e^- \to W^+
W^- $ process, obtaining limits based on the photon energy
distribution.
\end{abstract}

\section{Introduction} 
\label{int}

One of the main physics goals of LEP2 and future $e^+ e^-$
colliders is to directly test the gauge nature of couplings among
the electroweak gauge bosons. The process with largest cross
section at LEP2 involving these couplings is the $W$--pair
production, $e^+ e^- \to W^+ W^-$, which is sensitive to the
trilinear $WW\gamma$ and $WWZ$ couplings.  The measurement of
these couplings and the sensitivity to possible deviations from
the Standard Model (SM) predictions have been extensively studied
in the recent years \cite{reviews}.

The most general phenomenological parametrization for these
couplings \cite{param} can be achieved by means of an effective
Lagrangian \cite{effective} that involves  operators with
dimension higher than four, containing the relevant fields at low
energies and respecting the symmetries of the Standard Model.
The effective Lagrangian approach is a model--independent way to
describe  new physics that can occur at an energy scale $\Lambda$
much larger than the scale where the experiments  are performed.

The effective Lagrangian depends on the particle content at low
energies and since the Higgs boson has not yet been found, there
are two logical possibilities to describe the new physics effect
at low energies. In one of them, the Higgs boson can be light,
being present in the higher dimensional operators, in addition to
the electroweak gauge bosons, and the SM symmetries are linearly
realized \cite{linear,hisz}. Alternatively, the Higgs boson can
be very heavy and it must be integrated out at low energies. In
this case, the relevant fields at low energies are only
electroweak gauge bosons and the SM symmetries are realized
non-linearly \cite{nonlinear}.  Here we focus on a linearly
realized $SU_L(2) \times U_Y(1)$ invariant effective Lagrangian
to describe the bosonic sector of the Standard Model, keeping the
fermionic couplings unchanged.

The same effective Lagrangian used to describe anomalous
trilinear gauge couplings can, in general, lead to anomalous
quartic interaction among gauge bosons and also to anomalous
couplings of these particles with the Higgs field.   All these
interactions should also be investigated at LEP2 and at the Next
Linear Colliders (NLC)  in order to search for hints about the
nature of the new physics described by these higher dimensional
operators.

New quartic gauge boson couplings have been studied before in
many different processes at future $e^+ e^-$, $e \gamma$, $\gamma
\gamma$, $e^- e^-$ and $p p $ colliders \cite{quartic}. However,
most of these previous works have focused on the so--called
genuinely quartic operators, {\it i.e.}  operators that give rise
only to quartic gauge boson interactions without altering the
trilinear couplings \cite{genuine}.  Since these operators do not
appear in a dimension--6 linearly realized $SU(2)_L \times
U_Y(1)$ invariant effective Lagrangian \cite{foot}, they will not
be considered here. Anomalous Higgs boson couplings have also
been studied before in Higgs and Z boson decays \cite{hagiwara2},
in $e^+ e^-$ \cite{ee,our} and $\gamma\gamma$ colliders
\cite{gamma}.

The process with largest cross section in $e^+ e^-$ colliders
that also involves quartic couplings, and possibly anomalous
Higgs couplings, besides the trilinear couplings, is $e^+ e^- \to
W^+ W^- \gamma$.  Therefore, it is the most promising channel to
look for possible deviations from the Standard Model predictions.
This process has been considered by B\'elanger and Boudjema
\cite{genuine} and by Leil and Stirling \cite{stirling} in the
context of genuinely quartic operators, where the Higgs and
trilinear couplings were set to the Standard Model values and 
$3 \, \sigma$ deviations in the total cross section were used to
determine the reach of this reaction.  Grosse--Knetter and
Schildknecht \cite{grosse} have considered the effect of a {\it
single} higher dimensional operator usually denoted by ${\cal
O}_W$ (see below) in the above process, taking into account
modifications on both trilinear and quartic couplings.  However,
they assumed that the Higgs boson mass lies above the energy
region to be investigated and therefore they disregarded its
contribution. 

The purpose of this work is to study the sensitivity to these
anomalous couplings of the process $e^+ e^- \to W^+ W^- \gamma$
at LEP2 and the NLC.  We consistently include in our calculations
{\it all} new couplings introduced by the effective Lagrangian
that has become widely adopted to describe new physics beyond the
Standard Model.  In particular, this process is sensitive to
operators related to anomalous Higgs boson couplings that do {\it
not}  affect the self--coupling of gauge bosons and hence are not
constrained by the LEP2 measurements of $e^+ e^- \to W^+ W^- $.
Therefore, the process $e^+ e^- \to W^+ W^- \gamma$ may provide
important information about these operators at the NLC.

This paper is organized as follows. In Section \ref{eff:lag}, we
review the framework of effective Lagrangians that we use to
parametrize anomalous couplings and explain the methodology used
to study the $W^+ W^-\gamma$ production. In Section \ref{lep}, we
analyze the sensitivity at LEP2 based on the total cross section.
In Section \ref{nlc}, we study the improvements arising from
going to  NLC energies, the effects of having a polarized
electron beam, and the impact of being able to measure the $W$
boson polarization. We then concentrate on the analysis of
operators which will not be probed by  the $e^+ e^- \to W^+ W^-$
process, obtaining limits based on the photon energy spectrum. We
present our conclusions in Section \ref{con}.

\section{Effective Lagrangian and the Process
\lowercase{e}$^+$ \lowercase{e}$^-$ $\to W^+ W^- \gamma$ }
\label{eff:lag}

In order to write down the most general dimension--6 effective
Lagrangian containing all SM bosonic fields, {\it i.e.\/}
$\gamma$, $W^{\pm}$, $Z^0$, and $H$, we adopt the notation of
Hagiwara {\it et al.} \cite{hisz}. This Lagrangian has eleven
independent operators in the linear representation that are
locally $SU_L(2) \times U_Y(1)$ invariant, $C$ and $P$ even. We
discard the four operators which affect the gauge boson
two--point functions at tree--level and therefore are strongly
constrained by LEP1 measurements. We also do not consider the two
operators that modify only the Higgs boson self--interactions, 
since they  are not relevant for our calculations. We are then
left with five independent operators, and the Lagrangian is
written as,
\begin{equation}
{\cal L}_{\text{eff}} = {\cal L}_{\text{SM}} + \frac{1}{\Lambda^2} \left(  
f_{WWW} {\cal O}_{WWW} + f_{WW} {\cal O}_{WW} + f_{BB} {\cal O}_{BB} +
f_W {\cal O}_{W} +f_B {\cal O}_{B} \right)
\; , 
\label{lagrangian}
\end{equation}
with each operator ${\cal O}_i$ defined as, 
\begin{eqnarray}
{\cal O}_{WWW} &=& \text{Tr} 
\left[ \hat{W}_{\mu \nu} \hat{W}^{\nu \rho} 
                   \hat{W}_{\rho}^{\mu} \right] \\
{\cal O}_{WW} &=& \Phi^{\dagger} \hat{W}_{\mu \nu} 
\hat{W}^{\mu \nu} \Phi \\
{\cal O}_{BB} &=& \Phi^{\dagger} \hat{B}_{\mu \nu} 
\hat{B}^{\mu \nu} \Phi \\
{\cal O}_{W}  &=& (D_{\mu} \Phi)^{\dagger} 
\hat{W}^{\mu \nu} (D_{\nu} \Phi) \\
{\cal O}_{B}  &=& (D_{\mu} \Phi)^{\dagger} 
\hat{B}^{\mu \nu} (D_{\nu} \Phi) \; ,
\end{eqnarray}
where $\Phi$ is the Higgs field doublet, which in the unitary
gauge assumes the form,
\[
\Phi = \left(\begin{array}{c}
0 \\
(v + H)/\sqrt{2}
\end{array}
\right) \; , 
\]
and
\begin{equation}
 \hat{B}_{\mu \nu} = i \frac{g'}{2} B_{\mu \nu}  \;\; , \;\; \; 
\hat{W}_{\mu \nu} = i \frac{g}{2} \sigma^a W^a_{\mu \nu} \; , 
\label{lagr}
\end{equation}
with $B_{\mu \nu}$ and $ W^a_{\mu \nu}$ being the field strength
tensors of the $U(1)$ and $SU(2)$ gauge fields respectively.

The operator ${\cal O}_{WWW}$ contributes only to anomalous gauge
couplings, ${\cal O}_{WW}$ and ${\cal O}_{BB}$ contribute only to
anomalous Higgs  couplings, $H Z Z$ and $H Z \gamma$, whereas
${\cal O}_{W}$ and ${\cal O}_{B}$ give rise to both types of new
couplings. Therefore, the existence of anomalous trilinear gauge
couplings could be related to the anomalous quartic gauge
couplings and Higgs interaction, which are the subject of our
investigation. 

Studies of anomalous trilinear gauge boson couplings from
$W$--pair production will significantly constrain combinations of
the parameters $f_{WWW}$, $f_W$ and $f_B$. However they are
``blind" with respect to $f_{WW}$ and $f_{BB}$. We chose to study
the reaction $e^+ e^- \to W^+ W^- \gamma$ since it is the process
with the largest cross section involving triple, quartic gauge
boson  couplings and also anomalous Higgs--gauge boson couplings.
Therefore, it is also sensitive to $f_{WW}$ and $f_{BB}$,
offering an excellent possibility for a detailed study of these
couplings.

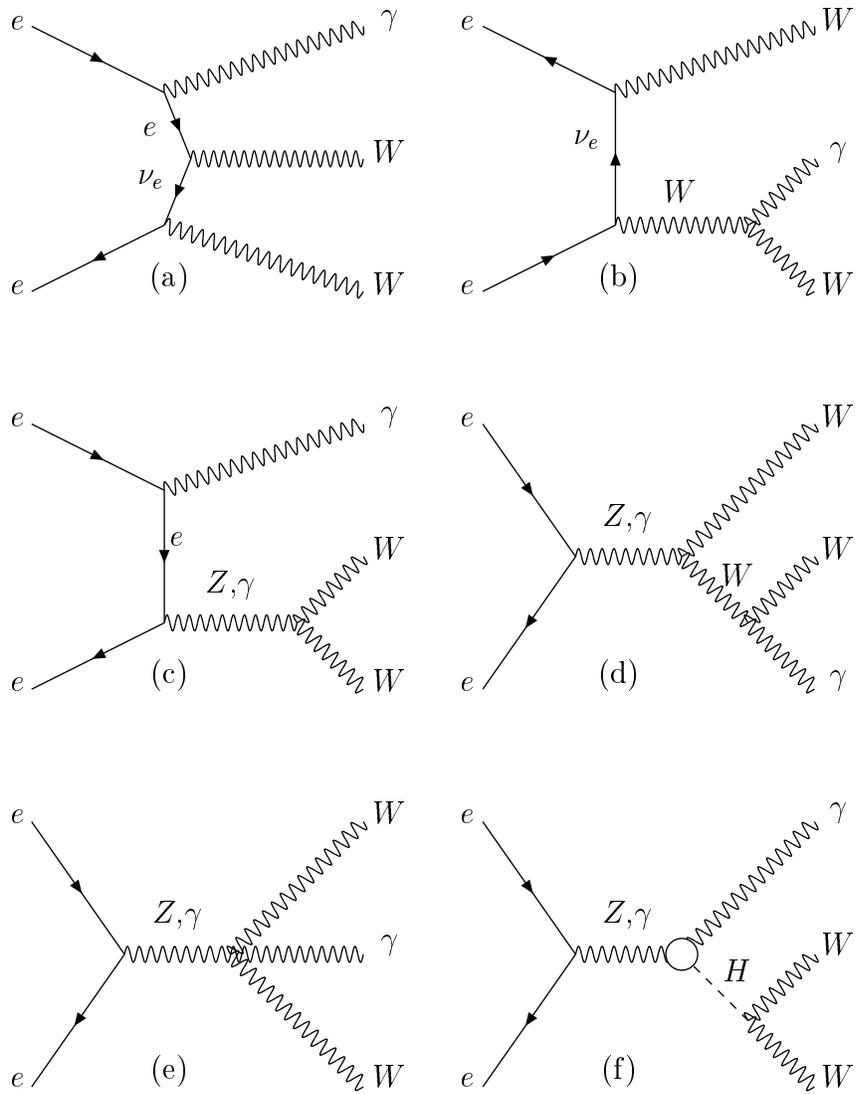
\begin{figure}
\protect
\caption{Feynmam diagrams for the Standard Model process  $e^+
e^- \rightarrow W^+ W^- \gamma$. Crossed diagrams are not shown.}
\label{fig:1}
\end{figure}

The Standard Model cross section for the process $e^+ e^- \to W^+
W^- \gamma$ was evaluated in Ref.\ \cite{SM}. When we neglect the
electron mass, Higgs contributions for this reaction do not
appear at tree level since the couplings $H \gamma \gamma$ and
the $HZ\gamma$ are generated only at one loop \cite{hgg,hgz}.
Taking into account these contributions, there are 16 Feynman
diagrams involved in the reaction $e^+ e^- \to W^+ W^- \gamma$,
which are represented in Fig.\ \ref{fig:1} (the crossed diagrams
are not shown) which yields,
\begin{eqnarray} 
\sigma_{WW\gamma}^{SM} &=  46 \;(418) \;\text{fb} , \; &
\; \text{with} \; E_\gamma > 20 \;(5)\;\text{GeV} , 
          \hspace{0.5cm}  \; \text{at} \;  \sqrt{s} = 190 \;
\text{GeV} \nonumber \\ 
\sigma_{WW\gamma}^{SM} &=  144 \;\text{fb} ,  &
\; \text{with} \; E_\gamma > 20 \; \text{GeV} , \hspace{1.1cm}
        \; \text{at} \; \sqrt{s} = 500 \;\text{GeV} 
\end{eqnarray} 
where we have required that the angle between any two particles
is larger than $15^\circ$. The cross section peaks at roughly
$\sqrt{s} = 300$ GeV and is typically two orders of magnitude
smaller than the two--body process $e^+ e^- \to W^+ W^-$, used to
constrain anomalous trilinear couplings.

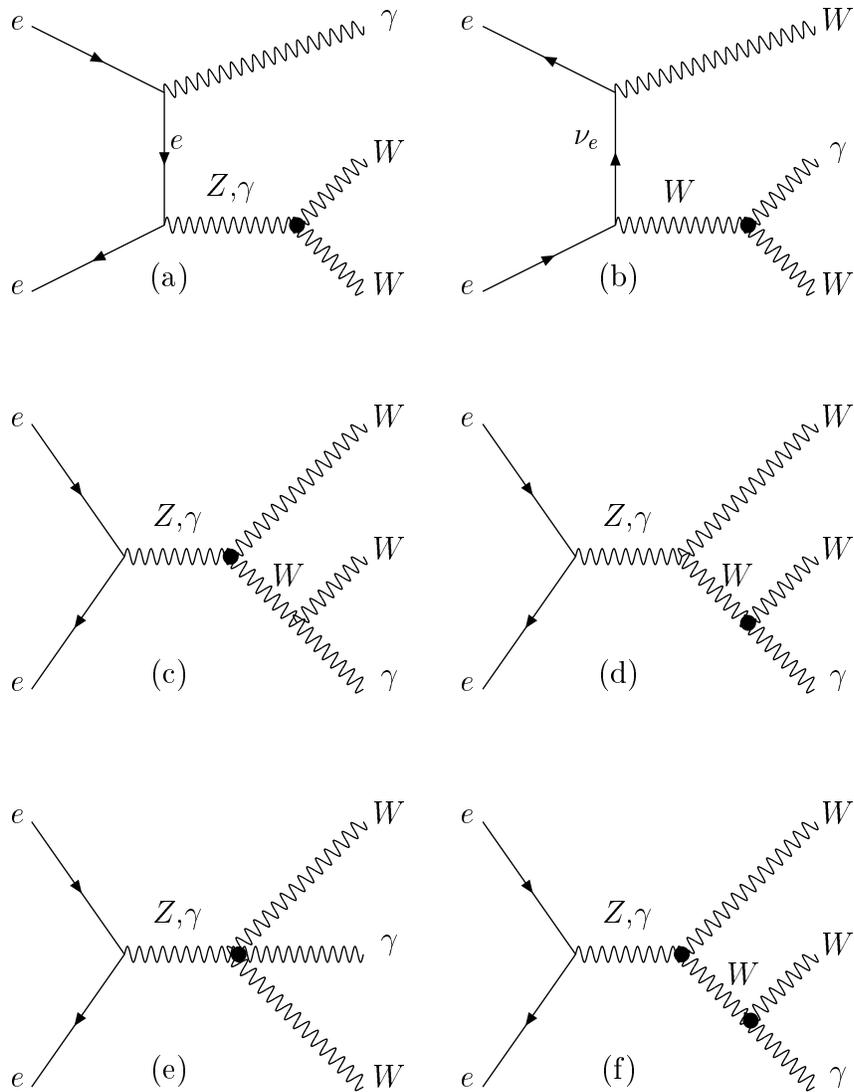
\begin{figure}
\protect
\caption{The vector bosons anomalous contributions to $e^+ e^-
\rightarrow W^+ W^- \gamma$. Crossed diagrams are not shown.}
\label{fig:2}
\end{figure}

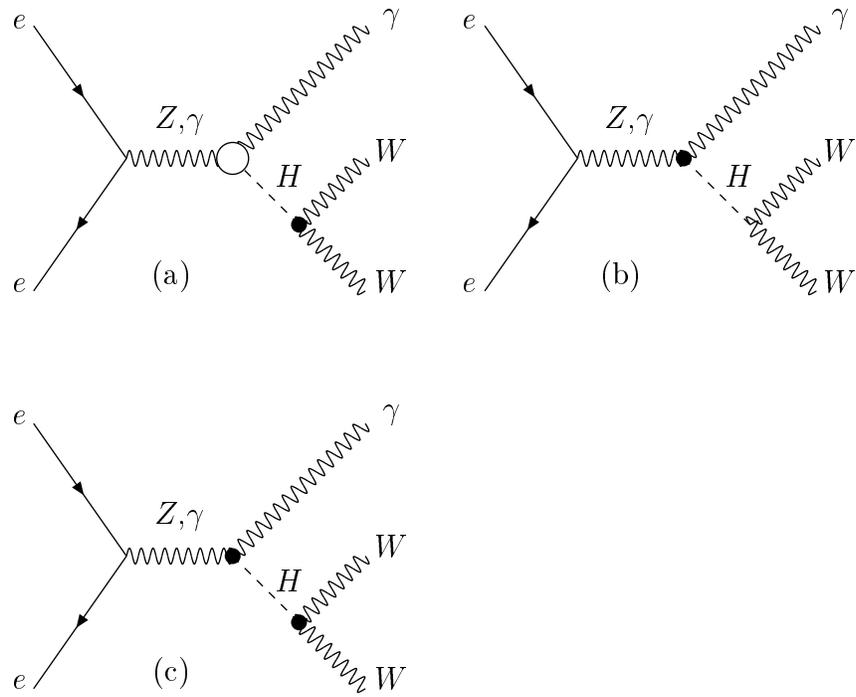
\begin{figure}
\protect
\caption{The Higgs boson anomalous contributions to $e^+ e^-
\rightarrow W^+ W^- \gamma$. }
\label{fig:3}
\end{figure}

In order to compute the contribution from all possible anomalous
couplings, we have developed a {\tt Mathematica} code to
automatically generate the Feynman rules for the Lagrangian
(\ref{lagrangian}) that were then incorporated in {\tt
Helas}--type \cite{helas} subroutines. These new subroutines were
used to extend a {\tt Madgraph} \cite{madgraph} generated code to
include all the anomalous contributions and to numerically
evaluate the helicity amplitudes and the squared matrix element.
In our calculations, we have taken into account the standard loop
Higgs contributions besides all the relevant anomalous couplings,
which give rise to the 42 contributions shown in Fig.\
\ref{fig:1}, \ref{fig:2}, and \ref{fig:3}.  We have checked that
our code passed the non--trivial test of electromagnetic gauge
invariance.  We employed {\tt Vegas} \cite{vegas} to perform the
Monte Carlo phase space integration with the appropriate cuts to
obtain the differential and total cross sections. Moreover, we
have studied the angular variables in order to find optimal cuts
to improve the anomalous contribution over the SM signal.

\section{$WW\gamma$ Production at LEP2}
\label{lep}

We studied the reaction $e^+ e^- \to W^+ W^- \gamma$ at LEP2
assuming a center--of--mass energy of $\sqrt{s} = 190 \;
\text{GeV}$ and an integrated luminosity of ${\cal L} = 0.5
\;\text{fb}^{-1}$. We applied a cut in the photon energy
($E_\gamma > 5 \; \text{GeV}$), and we required the angle between
any two particles to be larger than $\theta_{ij} > 15^\circ$.

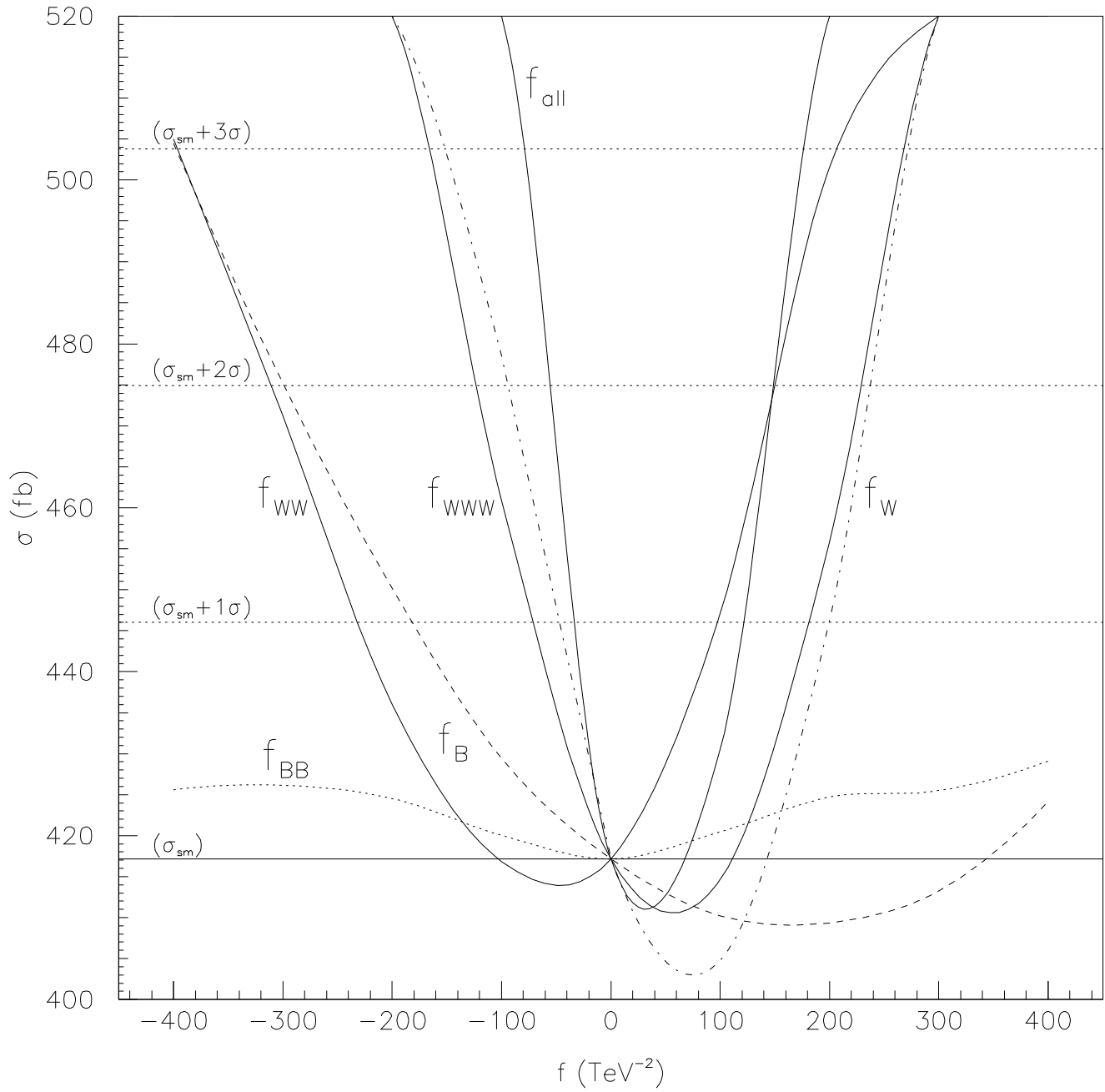
\begin{figure}
\protect
\caption{Total cross section (SM $+$ Anomalous) for the process
$e^+e^- \to W^+ W^- \gamma$, at LEP2 as a function of different
anomalous coefficients and also for the HISZ scenario
($f_{\text{all}}$). We assumed $m_H = 170$ GeV, and ${\cal L}$ =
0.5 fb$^{-1}$. The results for the SM and for 1, 2, and 3$\sigma$
deviations are displayed (see text for energy and angular cuts).}
\label{fig:4}
\end{figure}

Our results for the sensitivity of LEP2 to the operators
appearing in the effective Lagrangian (\ref{lagrangian}), from an
analyses of the total cross section, are summarized in Fig.\
\ref{fig:4} for a fixed value of the Higgs boson mass, $M_H =
170$ GeV.  We plot the contributions of the 5 different operators
separately, assuming that only one operator contributes each
time.  We also show the result for an extension of the so--called
HISZ scenario \cite{hisz}, where all the coefficient are
considered equal, {\it i.e.\/} $f_{WWW} = f_{WW} = f_{BB} = f_W =
f_B = f$, in order to reduce the number of free parameters to
only one ($f$). The Standard Model cross section and its value
with 1, 2 and 3 $\sigma$ deviations are depicted as horizontal
lines.

The most sensitive contribution comes from ${\cal O}_{WWW}$,
${\cal O}_{W}$, and ${\cal O}_{B}$. A 1$\sigma$ deviation in the
total cross section would be observed for the following ranges of
the coefficients of these operators, for $\Lambda = 1$ TeV,
\begin{equation}
-75 < f_{WWW} < 178  \;\;,\;\;  -48 < f_{W} < 192 \;\;,\;\; 
-188 < f_{B} < 550   \;\;,\;\;  -253 < f_{WW} < 110 \; ;
\label{bound:lep}
\end{equation} 
whereas for the extended HISZ scenario, we have,
\begin{equation}
 -33 < f <  119 \;.
\label{lep2}
\end{equation}

Of course, the operators that also give rise to changes in the 
triple vector boson couplings can also be constrained at LEP2
via the reaction $e^+ e^- \to W^+ W^-$. A recent analyses of
$W$--boson pair production based on a log--likelihood fit to a
five--fold differential cross section obtained the
1$\sigma$ limits \cite{hagiwara1}, $|f_{WWW}| < 10$, $|f_{W}| <
7.1$, and  $|f_{B}| < 46$. However, one should keep in mind that 
this reaction is insensitive to $f_{WW}$ and $f_{BB}$, and therefore
the study of the process $e^+ e^- \to W^+ W^-
\gamma$ can provide further information on these operators, as we 
show in this paper.


The contribution of the anomalous couplings involving only the
Higgs boson, {\it i.e.\/} $f_{WW}$ and $f_{BB}$ (see Fig.\
\ref{fig:3}), is dominated  by on--mass--shell Higgs production
with the subsequent $H \to W^+ W^-$ decay, 
\begin{equation}
\sigma (e^+ e^- \to W^+ W^- \gamma) \;\propto\; 
\sigma (e^+ e^- \to H \gamma) 
\; \times 
 \frac{\Gamma (H \to W^+ W^-)}{\Gamma (H \to \text{all})} .
\end{equation}

For large values of the operator coefficients, the total Higgs
width is dominated by the anomalous decay $H \to \gamma \gamma$
\cite{hagiwara2}, which is also proportional to $f_{WW}$ and
$f_{BB}$.  On the other hand, the anomalous  width $\Gamma(H \to
W^+ W^-)$ depends only on $f_{WW}$. Therefore, the
contribution from the anomalous coupling $f_{BB}$ is much less
sensitive than the contributions from the other operators since
$\sigma(e^+ e^- \to W^+ W^- \gamma)$ becomes almost independent
of this coefficient. Fortunately, this is not the case if one is 
sensitive to small values of the coefficients, as will occur at 
the NLC study in the next Section.

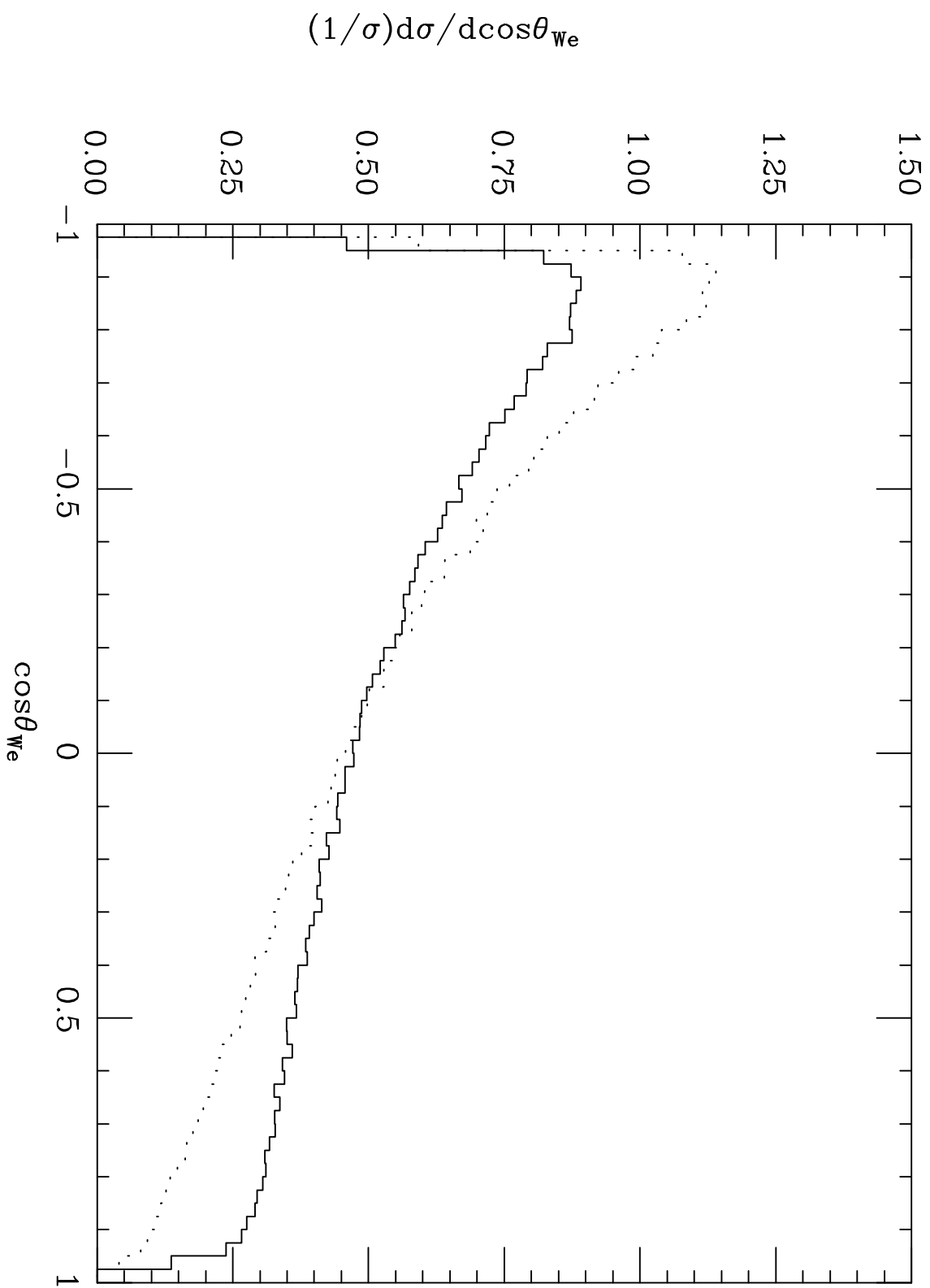
\begin{figure}
\protect
\caption{Normalized $W^{+}-e^+$ angular distribution. The solid
(dashed) line represents the SM (SM $+$ Anomalous) contribution for
$f_{\text{all}}/\Lambda^2 = 150$ TeV$^{-2}$ and $m_H = 170$ GeV.}
\label{fig:5}
\end{figure}

We have investigated various distributions to try to improve the
LEP2 sensitivity. The most promising distribution is the angular
distribution of the $W$ bosons with respect to the beam direction
(see Fig.\ \ref{fig:5}). We computed the total cross section with
the extra cut $\cos \theta_{W^+e^+} > 0$, as suggested by this
distribution, and found an increase in sensitivity from $2\sigma$
to $2.8 \sigma$. However, due to the small deviations in the
shape of the kinematical distributions and small statistics, no
further improvement seems to be possible. 

\section{$WW\gamma$ Production at NLC}
\label{nlc}

The effect of the anomalous operators becomes more evident with
the increase of energy, and we are able to put tighter
constraints on the coefficients by studying their contribution to
different processes at the Next Linear Collider. We studied the
sensitivity of NLC to the process $e^+ e^- \to W^+ W^- \gamma$
assuming $\sqrt{s} = 500 \; \text{GeV}$ and an integrated
luminosity  ${\cal L} = 50 \;\text{fb}^{-1}$. We adopted a cut in
the photon energy of $E_\gamma > 20 \; \text{GeV}$ and required
the angle between any two particles to be larger than $15^\circ$.
We have analyzed this process for different values of the Higgs
boson mass.  

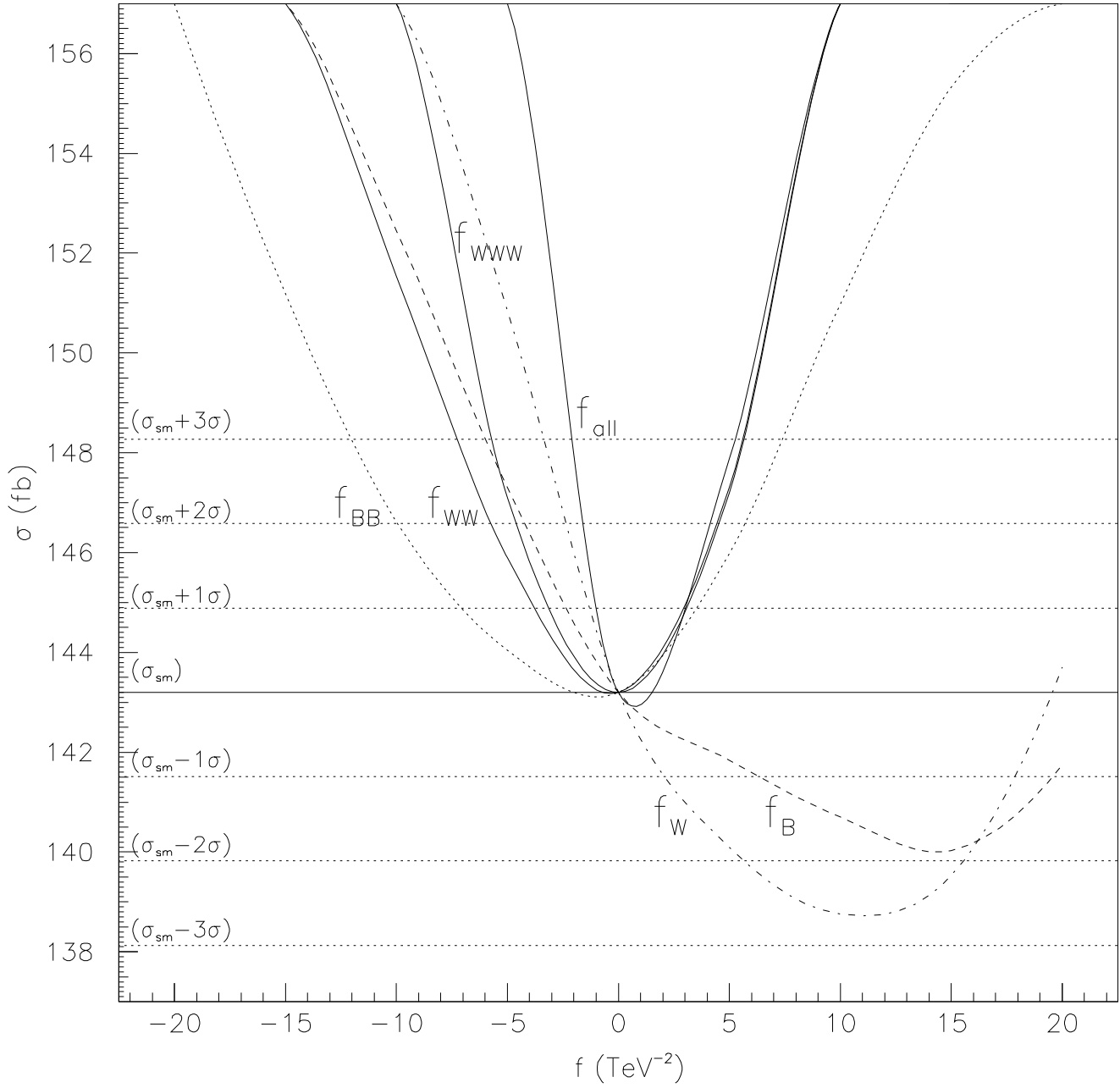
\begin{figure}
\protect
\caption{The same as Fig.\ \protect\ref{fig:4} for NLC, with
$\protect\sqrt{s}=500$  GeV and  ${\cal L}$ = 50
fb$^{-1}$.}
\label{fig:6}
\end{figure}

In Fig.\ \ref{fig:6}, we show the results for the total cross
section, for $M_H = 170 \; \text{GeV}$, including the effects of
the anomalous operators. The values of the coefficients $f$'s for
which a $2 \sigma$ deviation is obtained are shown in Table
\ref{tab:1}, being typically of the order of $1-10$ TeV$^{-2}$.
As we could expect, the $W$--pair production at NLC is able to
put a limit that is one order of magnitude better for the
coefficients $f_{B,W,WWW}$ \cite{hagiwara1}. However this latter
reaction is {\it not} able to constraint $f_{BB,WW}$. 

In an attempt to increase the sensitivity, we looked at the
effects of a $90\%$ polarized electron beam in order to reduce
the SM background, mainly the one coming from diagrams of Fig.\
\ref{fig:1}a and \ref{fig:1}b were just left--handed electrons
are present.  We have considered both left--handed (LH) and
right--handed (RH) polarizations, expecting a larger anomalous
sensitivity for RH electrons.

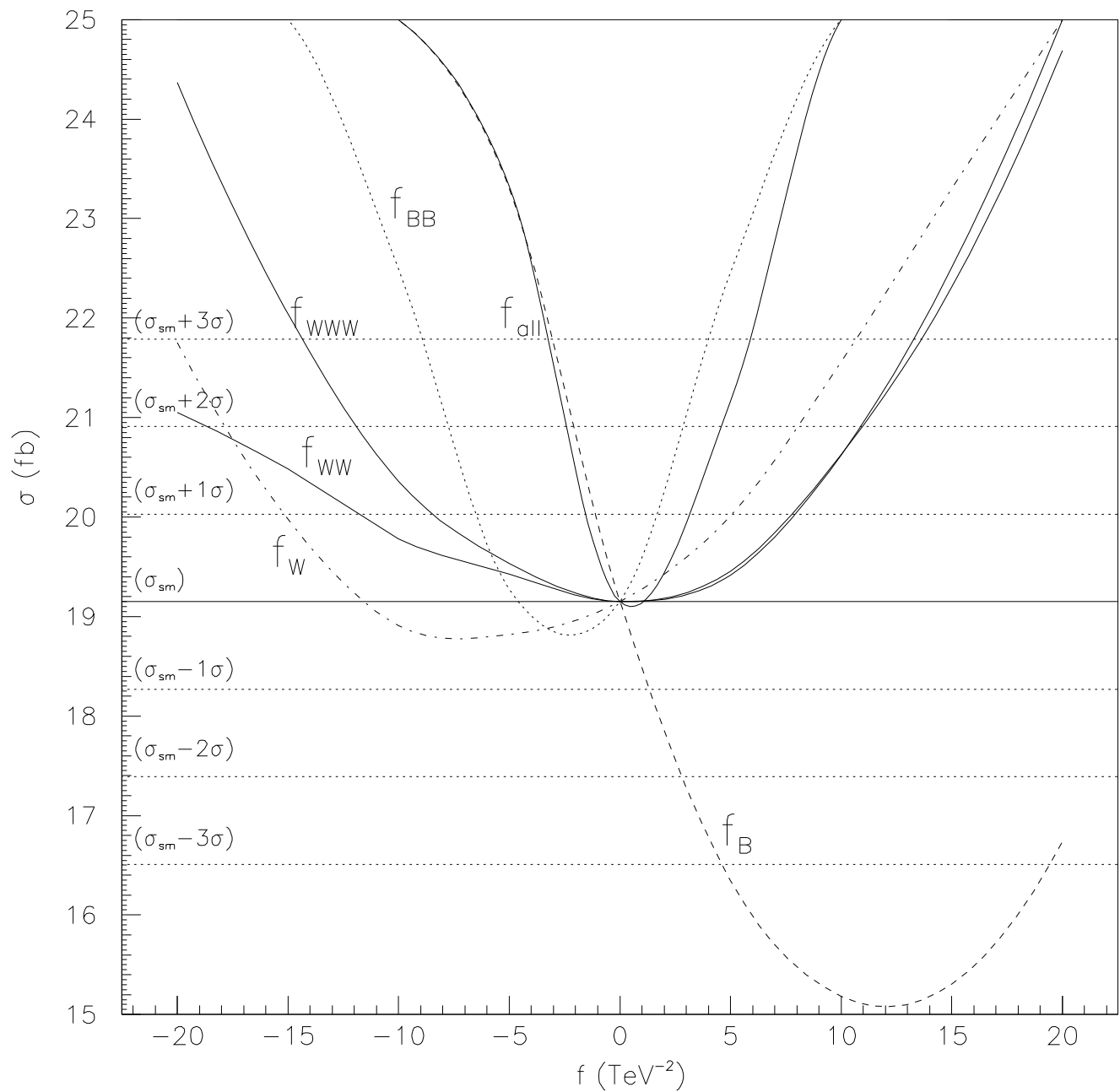
\begin{figure}
\protect
\caption{The same as Fig.\ \protect\ref{fig:6} for a $90\%$ 
right--handed polarized electron , with
$\protect\sqrt{s}=500$  GeV and  ${\cal L}$ = 50
fb$^{-1}$.}
\label{fig:7}
\end{figure}

In Fig.\ \ref{fig:7}, we show the results for the total cross
section, for a $90\%$ right--handed (RH) polarized electron, for
$M_H = 170$ GeV.  Comparing Fig.\ \ref{fig:6} and Fig.\
\ref{fig:7}, we notice that the effect of the anomalous
contributions in the total cross section are larger for the
polarized case. However, the small absolute value of the cross
section for the polarized case reduces the statistics and leads
to no improvement in the established limits, as shown in Table
\ref{tab:2}.

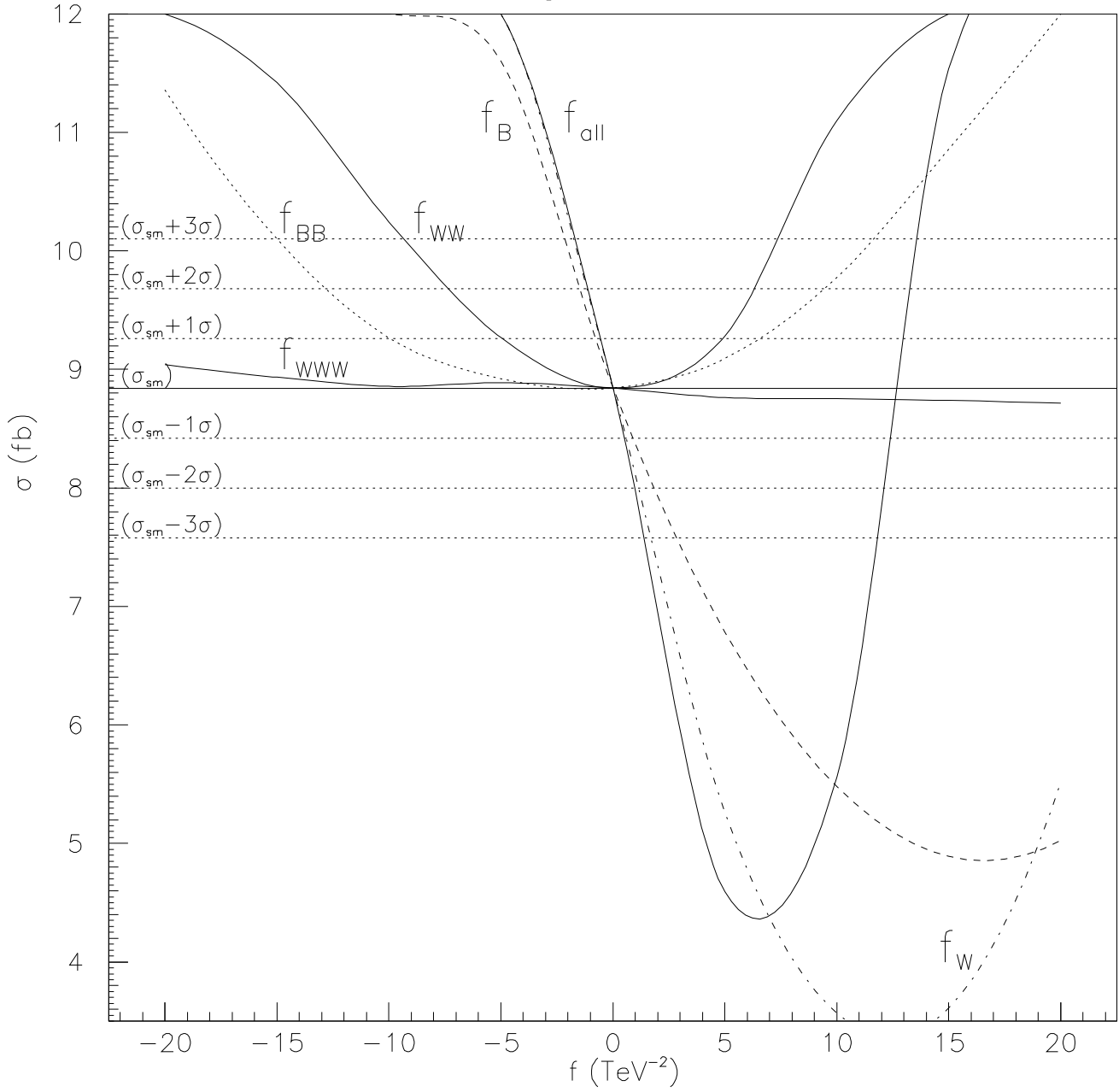
\begin{figure}
\protect
\caption{The same as Fig.\ \protect\ref{fig:6} for longitudinal
$W$ bosons ($W_L W_L$), with $\protect\sqrt{s}=500$  GeV and  
${\cal L}$ = 50 fb$^{-1}$.}
\label{fig:8}
\end{figure}

Since we expect the new interactions to involve mainly
longitudinally polarized gauge bosons, we studied the sensitivity
for different combinations of the polarizations of the $W-$pair.
In Fig.\ \ref{fig:8}, we show the analogous of Fig.\ \ref{fig:6}
for the $W_L W_L$ case. Again, the effect of the anomalous
contributions to the total cross section is increased, but no
further improvements are found due to the small statistics.  The
results for the bounds on the anomalous coefficients for the $W_L
W_L$, $W_T W_T$, and $(W_L W_T + W_T W_L)$ cases can be seen in
Table \ref{tab:3}. These bounds were obtained requiring a
2$\sigma$ effect on the total cross section.

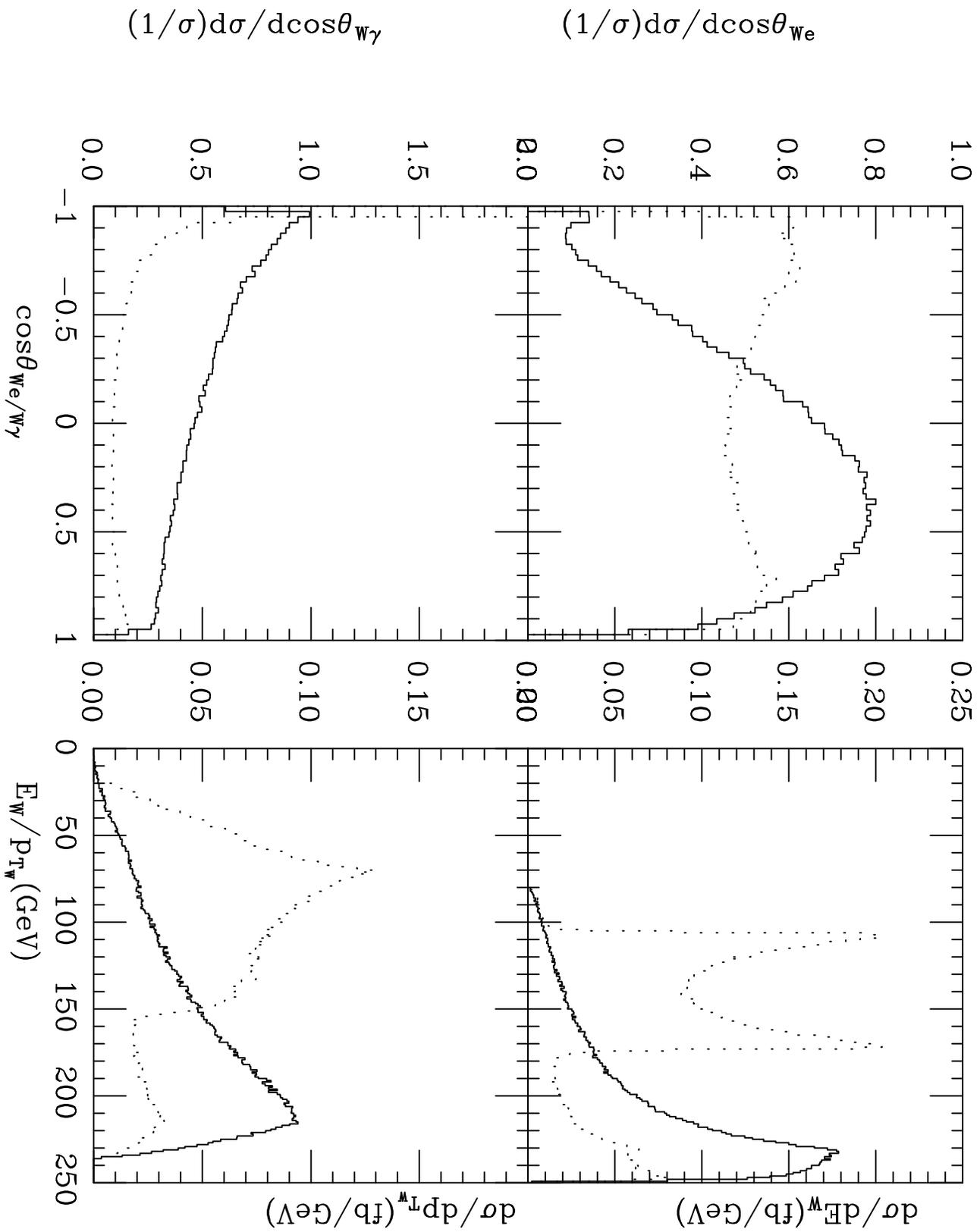
\begin{figure}
\protect
\caption{Kinematical distributions of the longitudinally
polarized $W^+$ vector boson for the SM (solid histogram) and for
the anomalous contribution (dotted histogram).}
\label{fig:9}
\end{figure}

It is important to notice that the kinematical distributions of
the longitudinally polarized $W$'s are quite different from the
SM results. As we could expect, the new physics effects becomes
more evident for longitudinal $W$'s since the decay $H \to W^+
W^-$ is dominated by this state of polarization.  In Fig.\
\ref{fig:9}, we present the angular distribution of the
longitudinal $W^+$ boson  with the initial positron and with the
final photon, the energy and the transverse momentum
distributions. We can see, for instance, that the $W$ energy
distribution is very different from the SM prediction. Its
characteristic behavior for $100 < E_W < 175$ GeV is due to the
presence of the Higgs boson, which decays into the $W$ pair
giving rise, at the same time, to a monochromatic photon. 

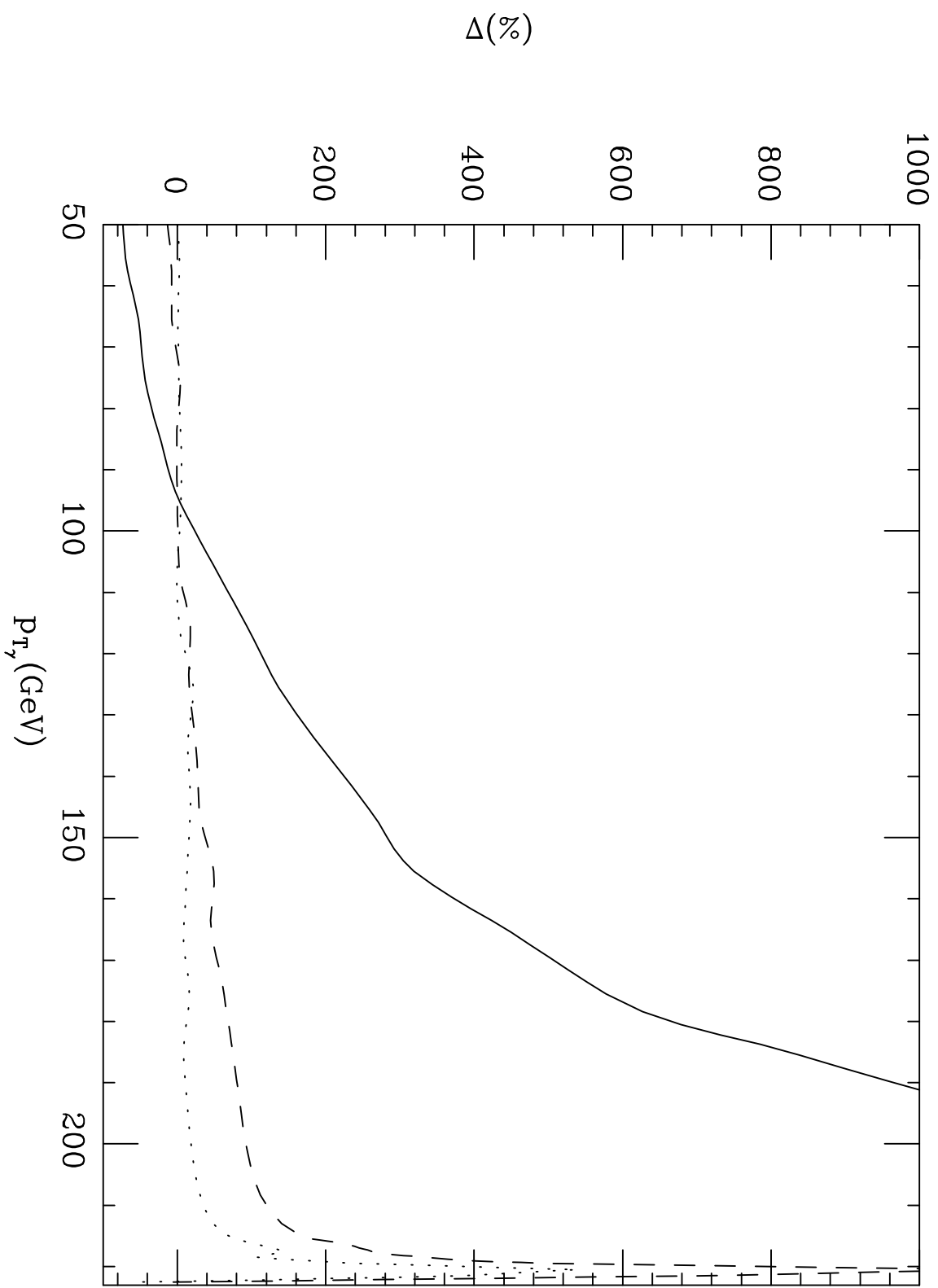
\begin{figure}
\protect
\caption{Plot of deviation ($\Delta$) in the photon
$P_{T_\gamma}$ distribution for the cases of $W_L W_L$ (solid
line), $W_L W_T + W_T W_L$ (dashed line) and $W_T W_T$(dotted
line).}
\label{fig:10}
\end{figure}

We present in Fig.\ \ref{fig:10} the percent deviation of the SM
prediction in the photon transverse momentum distribution, {\it
i.e.},
\[
\Delta = \left( \frac{d \sigma_{\text{ANO}}/dp_{T_\gamma} } 
{d \sigma_{\text{SM}}//dp_{T_\gamma} } - 1 \right ) \times 100 \% \; ,
\]
for the different polarization of the $W$'s. Once again the
relevance of the $W_L W_L$ case is evident: $\Delta > 100\%$ for
$p_{T_\gamma} > 120$ GeV. When a cut of $p_{T_\gamma} > 100$ GeV
is implemented, the background is drastically reduced and the
ratio of anomalous over SM events per year goes from $576/442$ to
$424/74$, for $f_{\text{all}} = 15$ TeV$^{-2}$. 

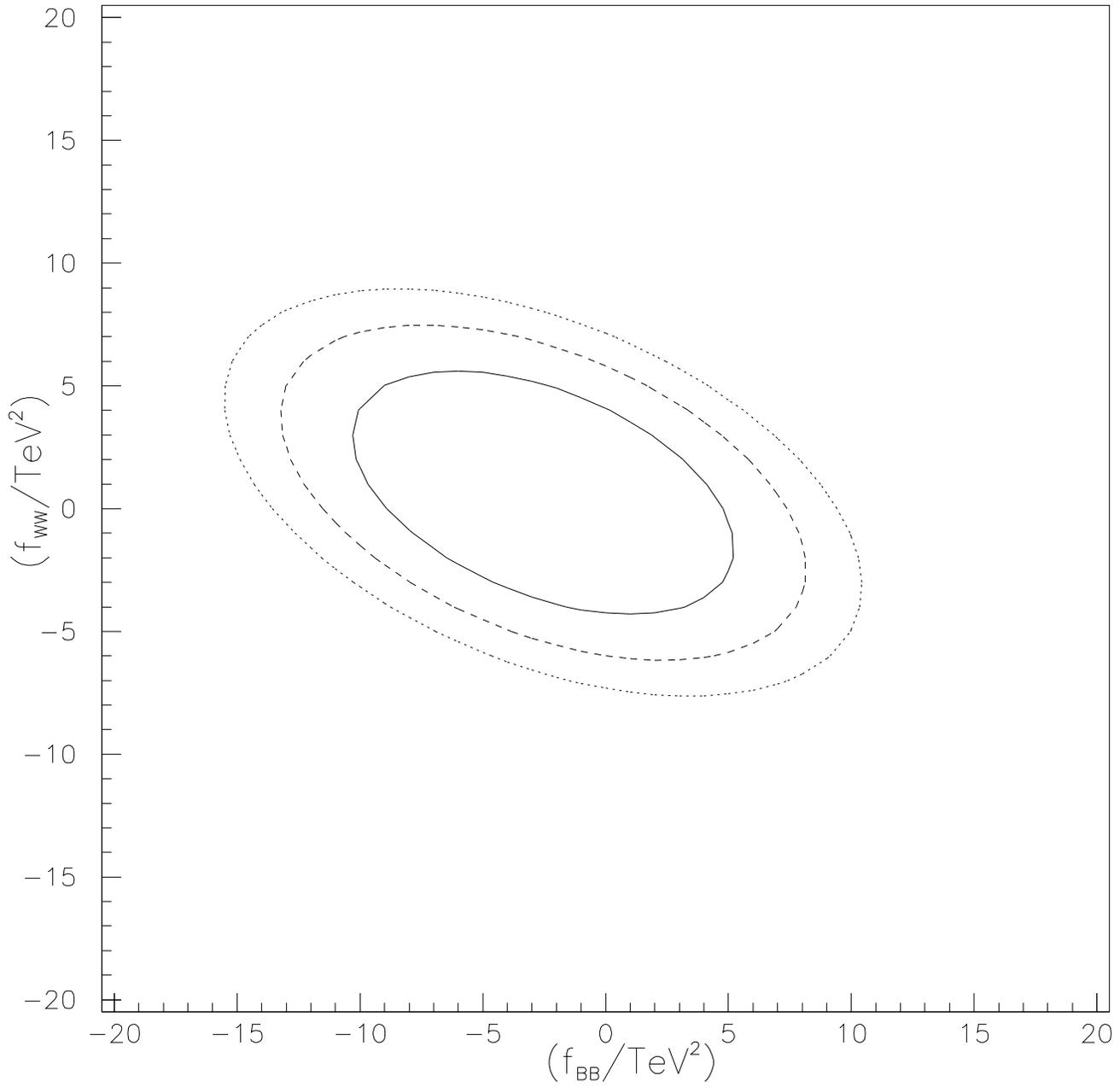
\begin{figure}
\protect
\caption{Contour plot of $f_{BB} \times f_{WW}$, for $M_H =170$
GeV. The curves show the one, two, and three sigma deviations 
from the Standard Model value of the total cross section.}
\label{fig:11}
\end{figure}

Using the reaction $e^+ e^- \to W^+ W^- \gamma$, we are also able
to establish bounds on the values of the coefficients $f_{WW}$
and $f_{BB}$, for which the $W$--pair process is insensitive,
since they only affect the Higgs boson couplings. In Fig.\
\ref{fig:11}, we present the results of a combined sensitivity
analysis in the form of a contour plot for the two free
parameter, $f_{BB}$ and $f_{WW}$, for $M_H = 170$ GeV. These are
the most relevant coefficients for the anomalous Higgs boson
phenomenology and they are not constrained by the $W-$pair
production. We should keep in mind that the $WW\gamma$ production
at LEP2 can put a 1$\sigma$ bound on $f_{WW}$ (\ref{bound:lep})
while it is not possible to impose a limit on $f_{BB}$ since the
cross section is quite insensitive to this coefficient.

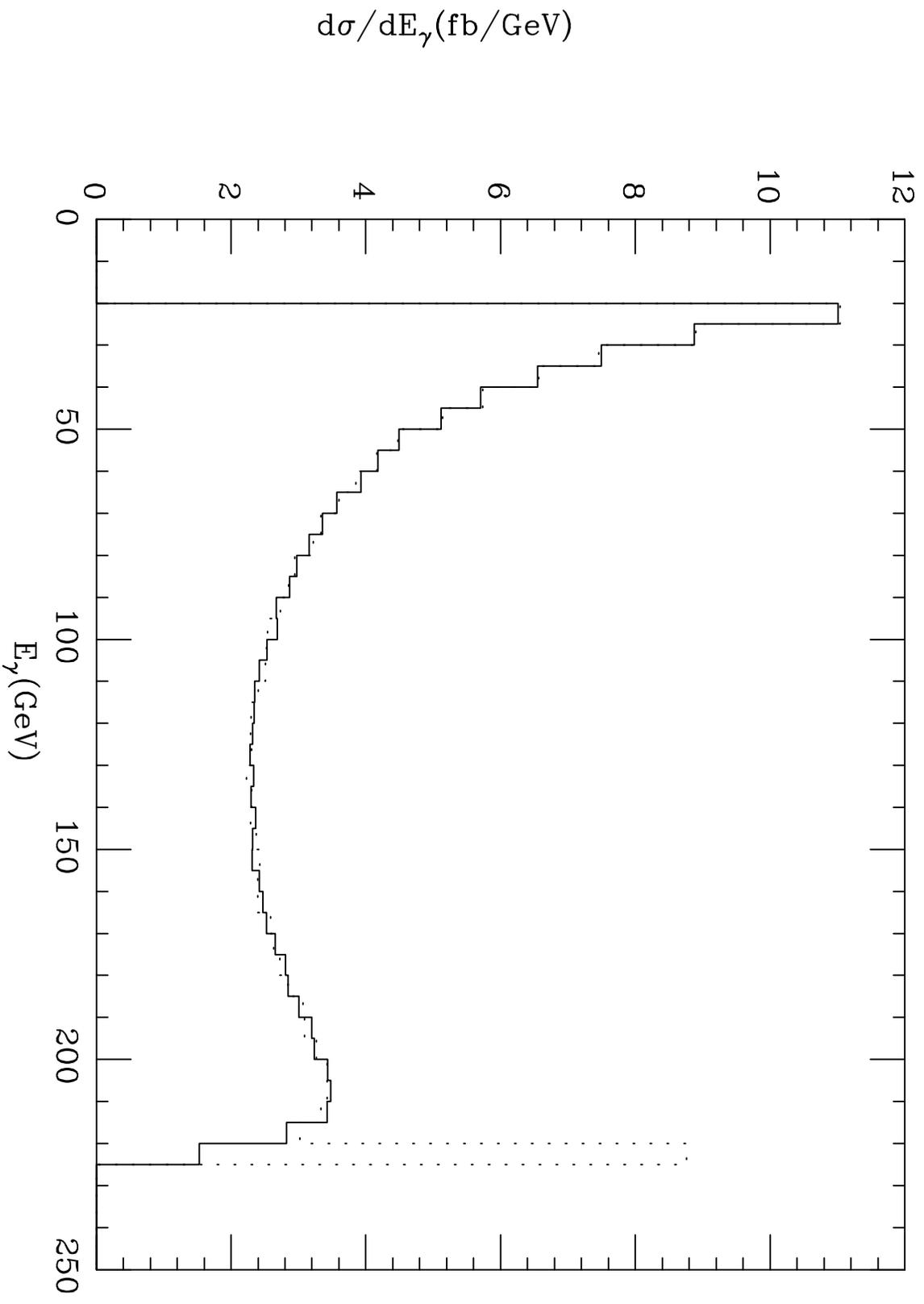
\begin{figure}
\protect
\caption{Photon energy distribution for the SM (solid line) and
for the SM $+$ Anomalous (dashed line), for $f_{WW}/\Lambda^2 = $
$f_{BB}/\Lambda^2 = 5 \; \text{TeV}^{-2}$, and $M_H =$ 170 GeV,
with a 5 GeV bin.}
\label{fig:12}
\end{figure}

If the Higgs boson is found with a mass in the range from 170 to
300 GeV, one would have a large sensitivity for the anomalous
Higgs couplings $f_{WW}$ and $f_{BB}$ in the photon energy
distribution of the process $e^+ e^- \to W^+ W^- \gamma$. This
increased sensitivity comes about because the existence of a peak
in the photon energy spectrum due to the $2$--body nature of the
dominant contribution, {\it i.e.\/}  $e^+ e^- \to H \gamma$
followed by the subsequent decay $H \to W^+ W^-$ (see Fig.\
\ref{fig:3}). In Fig.\ \ref{fig:12}, we illustrate this effect
with a typical photon energy distribution, for $f_{WW}/\Lambda^2
= f_{BB}/\Lambda^2  =  5$ TeV$^{-2}$ and $M_H = 170$ GeV, where
the Higgs peak appears very clearly in the photon spectrum of the
anomalous contribution.

In order to analyse the significance of the signal based  on the
photon energy spectrum, we took different energy bins of $1$, $3$
and $5$ GeV. The reason is to roughly mimic the effects of a
realistic  simulation including the finite energy resolution of
the detector and the small spread in the real center--of--mass
energy due to initial state radiation. We have not considered the
experimental efficiency, $\epsilon_{\text{eff}}$, for $W$
reconstruction. It can be easily incorporated  by multiplying the
obtained  significances by  $\sqrt{\epsilon_{\text{eff}}}$. Table
\ref{tab:4} shows the improvement on the sensitivity compared to
the total cross section analysis for the $f_{WW}/\Lambda^2 =
f_{BB}/\Lambda^2 = 5$ TeV$^{-2}$ and $M_H = 170 \;\text{GeV}$
case. 

In Table \ref{tab:5}, we present our  results for the sensitivity
on $f_{BB}/\Lambda^2$ and $f_{WW}/\Lambda^2$, assuming $f_{BB}$ =
$f_{WW}$, for the three energy bins above. We obtained a
sensitivity of the order of a TeV$^{-2}$ for $M_H = 170 \;
\text{GeV}$, decreasing by a factor of roughly four for  $M_H =
300 \; \text{GeV}$, which does not depend in a significant way of
the bin size. For larger Higgs boson masses, the cross section is
reduced due to phase space suppression. For smaller Higgs boson
masses, the cross section is reduced since the Higgs boson is
off--mass--shell, and in this case it would be better to study
processes like $e^+ e^- \to b \bar{b} \gamma$ or  $e^+ e^- \to
\gamma \gamma \gamma$ \cite{our}.

\section{Conclusion}
\label{con}

The search for the effect of higher dimensional operators that
give rise to anomalous bosonic couplings should be pursued
in all possible processes since the results may provide important
information on physics beyond the Standard Model. We have studied
here the production of a $W$--pair plus a photon in $e^+ e^-$
colliders in order to analyse  the contributions of anomalous
couplings arising from dimension--6 operators of a linearly
realized $SU_L(2) \times U_Y(1)$ invariant effective Lagrangian.
We have included {\it all} the anomalous trilinear and quartic
gauge couplings, as well as the anomalous Higgs couplings with
gauge bosons.

We present the  limits attainable at LEP2 and at NLC, including
the Standard Model irreducible background. Polarization of the
electron beam and of the  $W$--pair are found to be insufficient
to improve the limits obtained from the total cross section. 

We also focused on the operators ${\cal O}_{WW}$ and  ${\cal
O}_{BB}$, which cannot be tested in the  $W-$pair production
process. We showed, in particular, that for Higgs boson masses in
the range $M_H = 170 $--$300 \; \text{GeV}$, the photon energy
spectrum provides a sensitive signature for the anomalous Higgs
couplings. Typical sensitivities of a few TeV$^{-2}$ at the NLC
are obtained for these coefficients, providing complementary
information on different higher dimensional operators.

\acknowledgments
This work was supported by Conselho Nacional de Desenvolvimento
Cient\'{\i}fico e Tecnol\'ogico (CNPq), and by Funda\c{c}\~ao de
Amparo \`a Pesquisa do Estado de S\~ao Paulo (FAPESP).


\begin{table}
\begin{tabular}{||c||c||}
Anomalous Couplings & Unpolarized  \\
\hline 
\hline
$f_{\text{all}}/\Lambda^2$ & ( $-$2, 5 )  \\
\hline
$f_B/\Lambda^2$   & ( $-$5, 31 )  \\
\hline
$f_{BB}/\Lambda^2$  & ( $-$11, 7 )  \\
\hline
$f_W/\Lambda^2$   & ( $-$3, 23 )  \\
\hline
$f_{WW}/\Lambda^2$  & ( $-$8, 4 )  \\
\hline
$f_{WWW}/\Lambda^2$ & ( $-$5, 5 )  
\end{tabular}
\caption{The minimum and maximum values (min, max) of the
coefficients $f_i/\Lambda^2$ in units of TeV$^{-2}$ for a $2 \,
\sigma$ deviation of the unpolarized total cross section.}
\label{tab:1}
\end{table}

\begin{table}
\begin{tabular}{||c||c||c||}
Anomalous Couplings & $e^-_{LH}$ & $e^-_{RH}$ \\
\hline 
\hline
$f_{\text{all}}/\Lambda^2$    & ( $-$2, 5 ) & ( $-$3, 5 ) \\
\hline
$f_B/\Lambda^2$   & ( $-$11, 42 ) & ( $-$2, 26 ) \\
\hline
$f_{BB}/\Lambda^2$ & ( $-$19, 17 ) & ( $-$8, 3 ) \\
\hline
$f_W/\Lambda^2$   & ( $-$2, 26 ) & ( $-$17, 9 ) \\
\hline
$f_{WW}/\Lambda^2$  & ( $-$7, 5 ) & ( $-$15, 11 ) \\
\hline
$f_{WWW}/\Lambda^2$ & ( $-$5, 5 ) & ( $-$12, 11 )
\end{tabular}
\caption{The minimum and maximum values (min, max) of the
coefficients $f_i/\Lambda^2$ in units of TeV$^{-2}$ for a
$2\sigma$ deviation of the total cross section with 90\%
polarized LH and RH electrons for a reduced luminosity of ${\cal
L}$ = 25 fb$^{-1}$.}
\label{tab:2}
\end{table}
\begin{table}
\begin{tabular}{||c||c||c||c||}
Anomalous Couplings  & $W_L W_L$ & $W_T W_T$ & $(W_L W_T + W_T W_L)$ \\
\hline 
\hline
$f_{\text{all}}/\Lambda^2$  & ( $-$1, 14 ) & ( $-$5, 3 ) & ( $-$2, 7 ) \\
\hline
$f_B/\Lambda^2$    & ( $-$2, 35 ) & ( $-$29, 29 ) & ( $-$9, 36 ) \\
\hline
$f_{BB}/\Lambda^2$   & ( $-$12, 9 ) & ( $-$14, 10 ) & ( $-$13, 8 ) \\
\hline
$f_W/\Lambda^2$    & ( $-$1, 25 ) & ( $-$36, 22 ) & ( $-$4, 23 ) \\
\hline
$f_{WW}/\Lambda^2$   & ( $-$8, 6 ) & ( $-$9, 6 ) & ( $-$8, 5 ) \\
\hline
$f_{WWW}/\Lambda^2$  & ( $-$51, 96 ) & ( $-$6, 4 ) & ( $-$5, 25 ) 
\end{tabular}

\caption{The minimum and maximum values (min, max) of the
coefficients $f_i/\Lambda^2$ in units of TeV$^{-2}$ for a
$2\sigma$ deviation of the total cross section for different
combinations of the final state $W-$pair polarization.}
\label{tab:3}
\end{table}
\begin{table}
\begin{tabular}{||c||c||c||c||c||}
$M_H$ (GeV) & Total Cross Section & 1 GeV bin & 3 GeV bin & 5 GeV bin \\
\hline 
\hline 
170 & 4.2 & 52.2 & 43.1 & 35.8  \\
\hline 
\hline
200 & 2.8 & 17.8 & 20.9 & 17.8  \\
\hline 
\hline 
250 & 1.8 & 8.3 & 10.7 & 9.9  \\
\hline 
\hline 
300 & 1.0 & 2.5 & 3.8 & 4.3  
\end{tabular}

\caption{Number of standard deviations $\sigma$ from the Standard
Model from a sensitivity analysis based on the total cross
section compared to a sensitivity analysis based on the  peak of
the photon energy distribution,  considering a $1$, $3$ and $5
\text{GeV}$ bin for different values of the Higgs mass.  We fixed
$f_{WW}/\Lambda^2 = f_{BB}/\Lambda^2 = 5$ TeV$^{-2}$. }
\label{tab:4}
\end{table}
\begin{table}
\begin{tabular}{||c||c||c||c||c||}
$M_H$ (GeV) & Total Cross Section & 1 GeV bin & 3 GeV bin & 5 GeV bin \\
\hline 
\hline 
170 &($-$5.9, 2.8) & ($-$3.9, 0.3) & ($-$3.9, 0.4) & ($-$3.9, 0.5) \\
\hline 
\hline
200 &($-$6.4, 3.6) & ($-$4.4, 0.9) & ($-$4.1, 0.8) & ($-$4.2, 0.9) \\
\hline 
\hline 
250 &($-$7.0, 4.9) & ($-$4.2, 1.8) & ($-$3.9, 1.6) & ($-$4.0, 1.6) \\
\hline 
\hline 
300 &($-$8.3, 6.9) & ($-$6.2, 4.3) & ($-$5.1, 3.2) & ($-$4.9, 3.0)
\end{tabular}

\caption{The minimum and maximum values (min, max) of the
coefficients $f_i/\Lambda^2$ (for $f_{BB} = f_{WW} = f$) in units
of TeV$^{-2}$ that generate a $95 \%$ C.L.\ signal for the total
cross section analysis and for the photon energy spectrum
analysis with  1, 3 and 5  GeV energy bins for different values
of the Higgs mass. }
\label{tab:5}
\end{table}


\begin{references}
\bibitem{foot} Genuine quartic operators, without anomalous
trilinear couplings, are only possible in the non--linear
realization of the gauge symmetry for $ZZZZ$, $WWZZ$ and $WWWW$
couplings \cite{hagiwara1}.
\end{references}
\end{document}